# Design-Phase Buffer Allocation for Post-Silicon Clock Binning by Iterative Learning

Grace Li Zhang, Bing Li, Jinglan Liu, Yiyu Shi *Senior Member, IEEE* and Ulf Schlichtmann, *Member, IEEE*

*Abstract*—At submicron manufacturing technology nodes, process variations affect circuit performance significantly. To counter these variations, engineers are reserving more timing margin to maintain yield, leading to an unaffordable overdesign. Most of these margins, however, are wasted after manufacturing, because process variations cause only some chips to be really slow, while other chips can easily meet given timing specifications. To reduce this pessimism, we can reserve less timing margin and tune failed chips after manufacturing with clock buffers to make them meet timing specifications. With this post-silicon clock tuning, critical paths can be balanced with neighboring paths in each chip specifically to counter the effect of process variations. Consequently, chips with timing failures can be rescued and the yield can thus be improved. This is specially useful in high-performance designs, e.g., high-end CPUs, where clock binning makes chips with higher performance much more profitable. In this paper, we propose a method to determine where to insert post-silicon tuning buffers during the design phase to improve the overall profit with clock binning. This method learns the buffer locations with a Sobol sequence iteratively and reduces the buffer ranges afterwards with tuning concentration and buffer grouping. Experimental results demonstrate that the proposed method can achieve a profit improvement of about 14% on average and up to 26%, with only a small number of tuning buffers inserted into the circuit.

## I. INTRODUCTION

At advanced technology nodes, process variations have become relatively larger, and thus caused expensive overdesign due to timing margins reserved during the design phase. To meet the challenges imposed by process variations, previous methods model process variations as random variables and incorporate them into timing analysis directly, leading to a boom of research on statistical static timing analysis (SSTA) in the last decade [2]. With the knowledge of the

This work was partly supported by the German Research Foundation (DFG) as part of the Transregional Collaborative Research Centre "Invasive Computing" (SFB/TR 89).

A preliminary version of this paper was published as [1] in the Proceedings of the Design, Automation and Test in Europe (DATE) conference, 2016. The major improvement of this paper over [1] is to process multiple samples with an iterative learning procedure using a low-discrepancy sample sequence (Sobol sequence).

Grace Li Zhang, Bing Li, and Ulf Schlichtmann are with the Institute for Electronic Design Automation, Technical University of Munich (TUM), Munich 80333, Germany (e-mail: grace-li.zhang@tum.de; b.li@tum.de; ulf.schlichtmann@tum.de).

Jinglan Liu and Yiyu Shi are with the Department of Computer Science and Engineering, University of Notre Dame (e-mail: jliu16@nd.edu; yshi4@nd.edu).

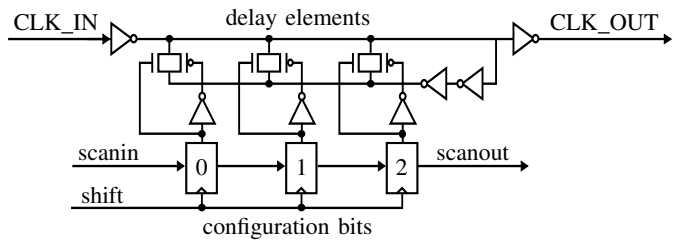

Fig. 1. Post-silicon tuning buffer in [4] with three configuration bits.

distributions of process variations, SSTA methods produce a performance-yield curve with which designers have a chance to make a tradeoff between different design goals. To alleviate the effect of process variations, many researchers have also worked on circuit structure level to introduce special devices and mechanisms. For instance, the Razor method [3] boosts circuit performance up to the limit where timing errors occur during circuit operation. Another technique to counter process variations is to use post-silicon tuning devices to adapt chips individually according to the effect of process variations after manufacturing.

A widely used post-silicon tuning technique is clock tuning with delay buffers. For example, the structure of the delay buffer used in [4] is illustrated in Fig. 1. The delay of this buffer can be changed by setting the configuration bits in the three registers. In high-performance designs, tuning buffers like this are inserted during the design phase. After manufacturing, the delay values of these buffers are configured to allot critical paths more timing budget by shifting clock edges toward the stages with smaller combinational delays. These critical paths might be different in individual chips due to process variations, so that only post-silicon tuning can counterbalance them efficiently. By balancing delay budgets across consecutive register stages, chips that might have failed to meet timing specifications can be revitalized, leading to an increased yield at the expense of additional area taken by these buffers. This post-silicon tuning technique works seamlessly with other optimization techniques, e.g., gate/wire sizing and timing-driven placement, since it mainly deals with delay imbalance introduced in manufacturing by process variations instead of during the design phase.

Post-silicon clock tuning buffers have various implementations and characteristics. The tuning buffer proposed in [5] provides precise adjustable delays of less than 30 ps by voltage-controlled driver strength. The design in [6] uses a delay line to generate delays with 1 ps resolution. The de-skew buffer in [7] consists of CMOS inverters and arrays of passive loads and is capable of creating a 170 ps tunable delay range in 8.5 ps steps. The controlled contention design [4] provides a 140 ps delay range with 8 steps. After manufacturing, these delays can be adjusted through the test access port (TAP) to tune individual chips.

In recent years, several methods have been proposed for statistical timing analysis and optimization of circuits with post-silicon clock tuning buffers. In [8] a clock scheduling method is developed and clock tuning buffers are selectively inserted to balance the skews due to process variations. In [9] algorithms are proposed to insert buffers into the clock tree to guarantee a given yield, while either the number of buffers or the total area of all buffers is minimized. The optimization problem is solved by evaluating the yield gradient with simultaneous perturbation and Monte Carlo simulation. In [10] the yield loss due to process variations and the total cost of clock tuning buffers are formulated together for gate sizing. The resulting optimization problem is solved using a stochastic cutting-plane method with an STA scheme based on Monte Carlo simulation. In [11], the placement of clock tuning buffers is investigated and a considerable benefit is observed when the clock tree is designed using the proposed tuning system. In addition, the work in [12] proposes an efficient post-silicon tuning method by searching a configuration tree combined with graph pruning, and an insertion algorithm to group buffers into clusters. The yield of a circuit with clock tuning buffers can be evaluated efficiently using the method in [13], and post-silicon testing methods for such circuits have been discussed in [14], [15].

The methods above are applied as pre-silicon optimization or post-silicon adjustment before shipping the manufactured chips to customers. Several other methods [16]–[18] have exploited these tuning buffers to improve circuit performance and reliability online, i.e., while the circuit is running. The method in [16] adjusts clock skews when the circuit is running according to timing errors to achieve a better performance in timing-speculative circuits. The method in [17] explores the insertion of clock tuning buffers and in-system configuration to reduce performance degradation due to aging. In addition, the method in [18] applies clock tuning buffers to compensate dynamic delay variations induced by temperature.

In order to take advantage of post-silicon tuning, these buffers should be inserted into the circuit during the design phase. Since they take die area and require special treatment during physical design, the number of tuning buffers in the circuit should be small to provide a good yield/profit improvement. This is essentially a statistical optimization problem when process variations are considered. Previous methods [9], [10] solve this problem by path search or the cutting plane method. In these methods, yield values of different combinations of buffer locations are evaluated using Monte Carlo simulation. New combinations of buffer locations are then selected to evaluate according to the yield gradient. This is in fact a statistical extension of linear programming. Since Monte Carlo simulation is used at many branching points, this direct extension requires a large runtime to determine buffer locations, though the calculated buffer locations may still fall into a local optimum in the problem space due to the nature of path search.

In this paper, we propose a method to determine buffer locations by iterative learning. In each iteration we try to capture the buffers that are important to the yield/profit of the circuit. Afterwards, we refine the identified buffer locations and compress buffer ranges to reduce area cost. The contributions of the proposed method are as follows:

- Instead of searching along a few paths in the problem space to find a set of buffer locations, we use representative sample points to identify the buffers that are important to the yield/profit directly. Using a low-discrepancy sample sequence, the proposed method can identify the proper buffer locations efficiently.
- We introduce a new way to model yield in representative samples to convert a statistical optimization problem into an ILP problem, so that heuristic statistical optimization can be avoided.
- We model the overall profit optimization problem instead of the yield at a given clock period. Consequently, the produced method can determine the buffer locations with respect to multiple clock bins in high-performance designs. When only one bin is used, this method is equivalent to the yield improvement problem with respect to a single clock period.
- The proposed sampling-based method produces tuning values in the representative chip samples. With these values, buffers can be grouped according to their tuning correlation to reduce area cost further.
- Compared with other methods, the proposed method is much faster, thanks to several acceleration techniques, even when the intermediate sample batches are not parallelized

The rest of this paper is organized as follows. We give an overview of timing constraints for circuits with post-silicon clock tuning buffers in Section II and formulate the buffer allocation problem in Section III. We explain the proposed method in detail in Section IV. Experimental results are shown in Section V. Conclusion and future work are given in Section VI.

## II. TIMING CONSTRAINTS WITH CLOCK BUFFERS

In a circuit with post-silicon tuning buffers, the delays of clock paths to flip-flops can be adjusted after manufacturing for each chip individually. The concept of this tuning can be explained using the example in Fig. 2a, where four flip-flops are connected into a loop by combinational paths. Without post-silicon clock tuning, the minimum clock period of this circuit is 8. If clock edges can be moved by adjusting the delays of these tuning buffers, the minimum clock period can be reduced to 5.5. For example, the buffer value $x_2$ shifts the launching clock edge at F2 0.5 units later and the buffer value $x_3$ shifts the launching clock edge at F3 3 units later. Therefore, with a clock period of 5.5, the combinational path between F2 and F3 now has 5.5-0.5+3=8 time units to finish signal propagation. This shifting of the clock edge reduces the timing budget of the path between F3 and F4 by 3 units, but this path still works with the clock period 5.5 because the buffer value $x_4$ moves the clock edge at F4 further later.

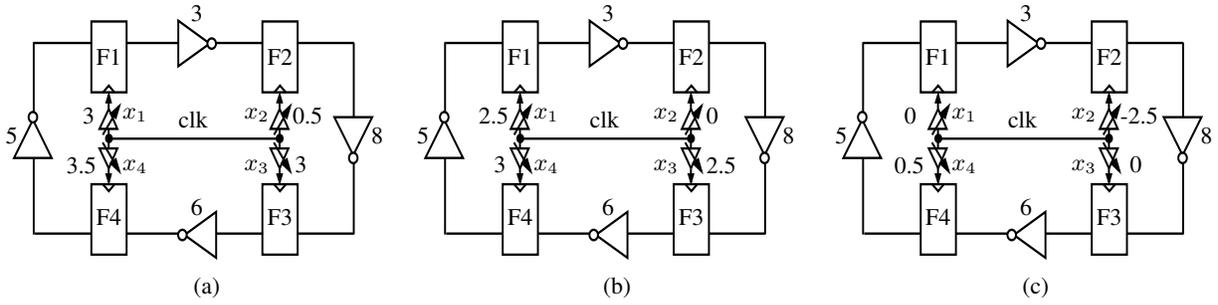

Fig. 2. Performance improvement using post-silicon tuning buffers. Minimum achievable clock period is 5.5. Tuning values in (a) and (b) are constrained in [0, 4]. Setup time and hold time are assumed as 0 for simplicity. (a) Tuning configuration without reduction. (b) Reduced tuning configuration. (c) Reduced tuning configuration with negative tuning values.

The timing imbalance between combinational paths as in Fig. 2a potentially appears when process variations become large in advanced technology nodes. For an individual chip, this post-silicon clock tuning is similar to the concept of useful clock skews [19]. The difference is that the tuning values are specific to each individual chip after manufacturing, so that the effects of process variations can be dealt with specifically for each chip. If the skew schedule problem in [19] is formulated with process variations, the skew to a flip-flop should still be identical in all manufactured chips, so that there is no chance to tune the chips with respect to the individual effect of process variations after manufacturing.

In Fig. 2a, four tuning buffers are used. However, all the delays of the buffers can be reduced by 0.5 time units and the circuit still works with the clock period 5.5. This way, the number of buffers can be reduced by one, as shown in Fig. 2b. Furthermore, we can reduce the number of buffers even to two, if we can move the clock edge at F2 2.5 time units earlier, so that the timing slack of the path between F1 and F2 can be shifted to the path between F2 and F3 directly, as in Fig. 2c. This negative delay can be implemented by shortening the original clock path in advance to introduce a negative delay in reference to the predefined arrival time of clock signals. With negative clock delays allowed, timing budgets can be balanced in both the clockwise direction and the counterclockwise direction, so that the number of required buffers can be lowered to reduce area and post-silicon configuration cost. The task of buffer allocation during the design phase is thus to identify the smallest set of buffers with which chips after manufacturing can be tuned to a higher performance.

The timing constraints with clock tuning buffers can be explained using Fig. 3, where two flip-flops with buffers are connected by a combinational circuit. Assume that the clock signal switches at reference time 0. Then the clock events at flip-flops $i$ and $j$ happen at time $x_i$ and $x_j$, respectively. To meet the setup time and hold time constraints, the following inequations must be satisfied.

$$x_i + \overline{d}_{ij} \leq x_j + T - s_j \quad (1)$$

$$x_i + \underline{d}_{ij} \geq x_j + h_j \quad (2)$$

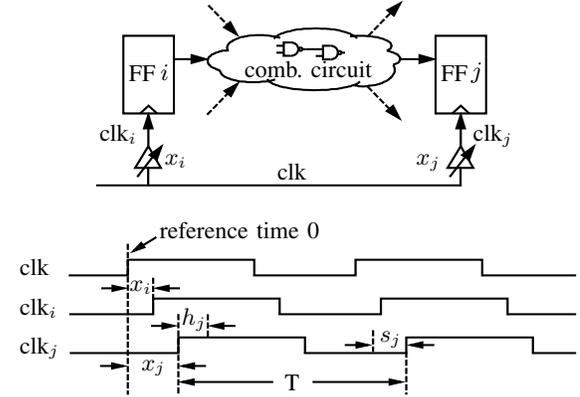

Fig. 3. Timing of circuits with tuning buffers.

where $x_i$ and $x_j$ are delay values of tuning buffers, $\overline{d}_{ij}$ ($\underline{d}_{ij}$) is the maximum (minimum) delay of the combinational circuit between flip-flops $i$ and $j$, $s_j$ ($h_j$) is the setup (hold) time of flip-flop $j$, and $T$ is the clock period. Here the clock buffers introduce two delay variables into the constraints (1) and (2). Without them, the two inequations fall back to the normal timing constraints of digital circuits.

Owing to area constraints, the configurable delay of a clock buffer usually has a limited range. Assume that the lower bound of the tuning values of buffer $i$ is $r_i$ and the upper bound is $r_i + \tau_i$, where $\tau_i$ is the size of the buffer. The delay value of buffer $i$ can thus be constrained by a range window as

$$r_i \leq x_i \leq r_i + \tau_i. \quad (3)$$

Unlike [9], we model the range window of the tuning values as asymmetrical with respect to 0 to achieve a maximal flexibility. Furthermore, $x_i$ may take only discrete values due to implementation limitations.

To guarantee the proper function of a circuit with clock tuning buffers, the constraints (1), (2) and (3) are created for each pair of flip-flops. For a chip after manufacturing, the variables $\overline{d}_{ij}$, $\underline{d}_{ij}$, $s_j$ and $h_j$ in (1) and (2) become fixed values. These delays and timing properties in manufactured chips can be measured using frequency stepping [20], such as in [14], [15], [21]. A detailed discussion of this technique can be found, e.g., in [21]. After the delays and timing properties are measured, the values of $x_i$ and $x_j$ that make a chip work with a given clock period $T$ can be found easily using the Bellman-Ford algorithm [22] or using linear programming.

In this paper, we only focus on determining buffer locations during the design phase. Consequently, the path delays, setup times and hold times should be considered as random variables modeled using data provided by foundries. With process variations considered, the tuning delays $x_i$ and $x_j$ also become statistical, because the clock buffers are subject to process variations too. These variations can be decomposed and merged with the random variables representing $\overline{d}_{ij}$, $\underline{d}_{ij}$, $s_j$ and $h_j$, e.g., using the canonical form in [23]. For convenience, we assume that a tuning delay can be configured to a fixed value in the following discussion. The task of buffer allocation is thus to determine the locations of buffers that can make as many chips as possible meet the given timing specification after manufacturing, using only the statistical timing information available during the design phase.

## III. PROBLEM FORMULATION

In applying the post-silicon tuning technique, we need to insert the buffers after logic synthesis is finished and before physical design is started. Since buffers take precious die area, and require additional test to configure them, the number of buffers in a design should be limited. In addition, the ranges of the buffers should be reduced as much as possible. Furthermore, in high-performance designs such as CPUs, chips are tested after manufacturing and assigned into bins of different performance grades, and the price of a chip from a bin of high speed is higher than that from a low-speed bin. In this scenario, it is more important to improve the overall profit of all bins than to improve the yield of the circuit with respect to a single clock period.

The important notations that appear in this paper are listed in Table I, and the problem of buffer allocation is formulated as follows.

**Input:**
- Circuit structure and statistical path delays;
- Buffer specification, including the maximum allowed size $\tau_i$ of buffers defined in (3) and the number of discrete steps in the tunable delay range;
- The maximum number of buffers allowed in the circuit $N_b$;
- The number of performance bins $N_p$. For the $m$th bin, an upper bound $T_{m,u}$ and a lower bound $T_{m,l}$ are defined by the designer. After manufacturing, a chip with a clock period $T$ assigned to the $m$th bin should meet $T_{m,l} < T \leq T_{m,u}$. For a chip in the $m$th bin, the average profit is given as $p_m$. For convenience, we order the bins from high performance to low performance, so that $T_{m,u} = T_{m+1,l}$.

**Output:**
- A set of flip-flops at which tuning buffers should be inserted on the their clock paths;
- The sizes of the buffers inserted into the circuit. These sizes must be no larger than the given maximum size $\tau_i$.

**Constraints:**

TABLE I
NOTATIONS

| | |
|---|---|
| $r_i, \tau_i$ | Lower bound and size of a buffer range |
| $N_b$ | Upper bound of the number of buffers |
| $N_p$ | Number of performance bins |
| $T_{m,l}, T_{m,u}$ | Lower and upper bounds of the $m$th bin |
| $p_m$ | Average profit of a chip in the $m$th bin |
| $y_m$ | Percentage of chips in the $m$th bin |
| $\mathcal{P}$ | Overall profit |
| $x_i^k, x_j^k$ | Tuning values for the $k$th sample |
| $\overline{d}_{ij}^k, \underline{d}_{ij}^k$ | Sampled delays |
| $s_j^k, h_j^k$ | Sampled setup time and hold time |
| $T^k$ | Clock period of the $k$th sample |
| $c_i$ | 0-1 variable indicating whether the $i$th flip-flop has a buffer |
| $b_m^k$ | 0-1 variable indicating whether the $k$th sample is assigned to the $m$th bin |
| $g_m^k$ | Auxiliary 0-1 variable to express $b_m^k$ |
| $N_s$ | Number of samples in the low-discrepancy sequence |
| $N_f$ | Number of samples in the low-discrepancy sequence after prefiltering |
| $N_t$ | Number of samples in one batch processed together |
| $\mathcal{B}, \mathcal{B}'$ | Buffer sets saving the allocation candidates |

- For any pair of flip-flops $i$ and $j$ with combinational paths between them, the constraints (1)–(3) hold;
- The number of buffers inserted in the circuit must not exceed $N_b$.

**Objectives:**
- Maximize the overall profit

$$\mathcal{P} = \sum_{m=1}^{N_p} p_m y_m \quad (4)$$

where $y_m$ is the percentage of the chips that are assigned into the $m$th bin after manufacturing;

- Reduce the sizes of the inserted buffers while maintaining the overall profit $\mathcal{P}$.

In the definition of bins, the first bin has the highest performance, and it has no lower bound for the clock period $T$, so that $T_{1,l}$ can be set to any value no larger than zero. After manufacturing, if a chip cannot be assigned to any of those bins, i.e., $T > T_{N_p,u}$, this chip is considered as a part of yield loss. The definition (4) is very general. If only one bin is used, this problem falls back to the yield improvement problem with respect to a single clock period.

In the problem formulation above, we do not include the number of tuning buffers as a part of the optimization objective, because the relation between profit and the number of buffers is very complex. With our formulation, designers can generate several combinations of buffer number and profit, and select the most appropriate setting according to their own cost model. If necessary, however, the number of buffers can also be moved from a constraint into the optimization objective (4), and the proposed method can still work with only a slight modification.

The predominant challenge in solving the optimization problem above comes from the random variables in (1) and (2), because only statistical timing information are available during the design phase when the buffers are allocated. The profit in (4) is thus defined similar to an expected value, which is slightly larger than the actual profit after manufacturing because statistical delays and timing properties cannot be measured exactly [21]. Another challenge is that the variables $x_i$ and $x_j$ in (1) and (2) may take only discrete values in the range window defined by (3). For example, the de-skew buffer in [7] can be configured to only 20 discrete delays. In this case, integer linear programming (ILP) becomes almost the only method available to deal with the constraint set defined by (1)–(3) after the random variables are fixed by sampling.

To deal with the large number of samples in the problem space, learning-based methods have been applied in the design automation field extensively, e.g., for statistical path selection considering large process variations [24], for sensor placement in dynamic noise management systems [25], and for parametric yield estimation for analog/mixed signal circuits [26]. In the following section, we will introduce an efficient iterative learning-based method to capture buffer locations for yield improvement.

## IV. BUFFER ALLOCATION USING REPRESENTATIVE SAMPLING

The buffer allocation problem is essentially a statistical optimization problem. In the linear constraints in (1)–(3) the path delays, setup times and hold times are correlated random variables. Instead of using path search or the cutting plane method as in previous methods, we solve this problem using statistical sampling. The basic idea is that we use a set of representative samples and model the numbers of samples in the different performance bins directly. We then determine buffer locations by maximizing the overall profit calculated from the yield values of these bins and the profit per chip for each bin. By sampling the random variables directly we can transform the statistical optimization problem into an ILP problem. Therefore, the relation between the statistical variables and the profit of the circuit can be established directly. With this relation, we can then capture buffer locations that are sensitive to yield/profit.

The flow of the proposed method is illustrated in Fig. 4. In this flow, we first generate a low-discrepancy sample sequence (Sobol sequence) and filter out the samples that are not affected by any buffers. Thereafter, we try to capture buffer locations and refine them iteratively. The ranges of buffers are compressed and the number of buffers is reduced by grouping in the end to reduce area cost. This flow will be explained in detail in the following sections.

### A. Sampling-based ILP modeling between statistical delays and profit

Consider the case that we generate $N_s$ samples from the joint distribution of all the random variables in the optimization problem. If $N_s$ is large enough, these samples can actually emulate the chips after manufacturing. If we have tuning buffers at the clock paths to some flip-flops, we can introduce intentional clock skews customized for each sample, or emulated chip, individually, to make the failing samples work again, or to move low-performance samples into high-performance bins. For each emulated chip we can now evaluate how performance can be improved because the statistical variables in the constraints become fixed in the samples. This way, we can establish the relation between buffer locations and the profit, and use an ILP solver to determine the optimal buffer allocation.

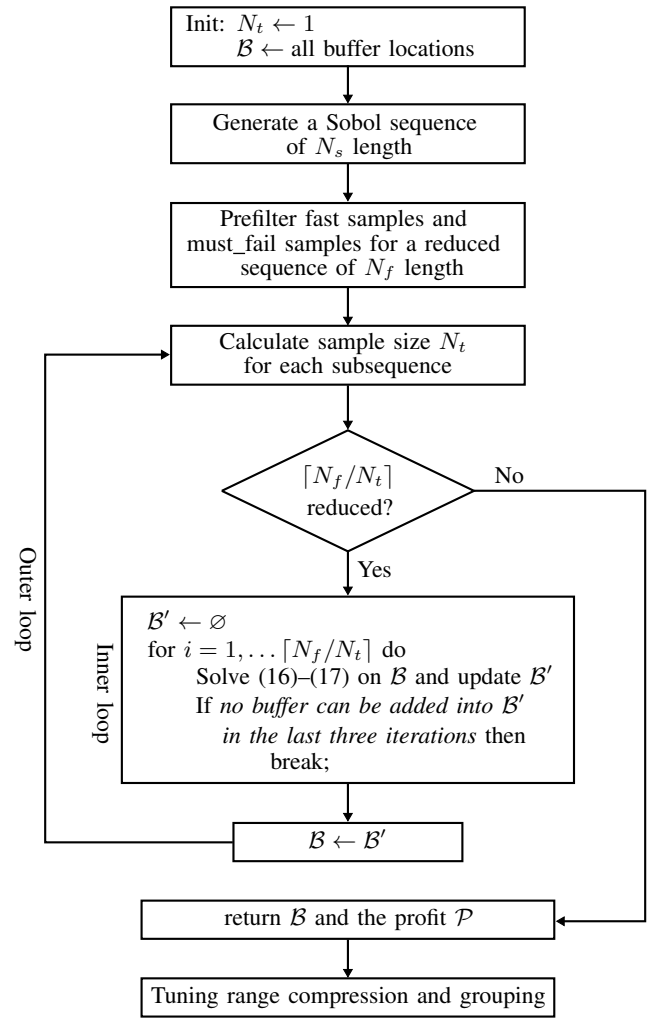

Fig. 4. Prefiltering and iterative buffer allocation flow.

For the $k$th sample from the $N_s$ samples, the constraints (1)–(3) become

$$x_i^k + \overline{d}_{ij}^k \leq x_j^k + T^k - s_j^k \tag{5}$$
$$x_i^k + \underline{d}_{ij}^k \geq x_j^k + h_j^k \tag{6}$$
$$r_i \leq x_i^k \leq r_i + \tau_i \tag{7}$$

where $\overline{d}_{ij}^k, \underline{d}_{ij}^k, s_j^k$ and $h_j^k$ are the $k$th sample values of random variables $\overline{d}_{ij}, \underline{d}_{ij}, s_j$ and $h_j$; $x_i^k$ and $x_j^k$ are intentional clock skews for this specific sample introduced by configuring the corresponding tuning buffers after manufacturing to improve the performance, in other words, to reduce the minimum clock period $T^k$. Note in (7) $r_i$ and $\tau_i$ are not indexed by $k$, because if a buffer is inserted on the clock path to a flip-flop, it appears in all the chips after manufacturing, and its range in all chips is also the same.

To indicate whether there is a buffer inserted on the clock path to the $i$th flip-flop, we assign a binary variable $c_i$ to it. If there is no buffer inserted, $c_i$ is set to 0; otherwise, $c_i$ is set to 1. Because a post-silicon clock skew can be added only when a buffer appears, the skew or the tuning value of the buffer at the $i$th flip-flop can be written as

$$x_i^k = \begin{cases} 0 & \text{if } c_i = 0, \\ \text{any valve} \in [r_i, r_i + \tau_i] & \text{when } c_i = 1. \end{cases} \quad (8)$$

According to the definition of $c_i$, we need only to force $x_i^k$ to be 0 to disable the potential clock tuning when $c_i$ is equal to 0. The constraint (8) can thus be transformed to

$$x_i^k \leq c_i \Gamma \quad (9)$$
$$-x_i^k \leq c_i \Gamma \quad (10)$$

where $\Gamma$ is very large constant. If $c_i$ is set to 0, $x_i^k$ must be set to 0 to meet (9) and (10). If $c_i$ is set to 1, these two constraints have no effect because $\Gamma$ is a predetermined constant larger than any possible value of $x_i^k$ or $-x_i^k$. In this case, $x_i^k$ is actually constrained by (7).

With $c_i$ defined to indicate the appearance of a buffer at the $i$th flip-flop, we can constrain the number of buffers in the circuit easily as

$$\sum_i c_i \leq N_b \quad (11)$$

where the sum on the left adds the $c_i$ variables for all flip-flops in the circuit together, and $N_b$ is the given upper bound of the number of buffers allowed in the circuit.

To evaluate the performance of an emulated chip, we need to compare the minimum clock period $T^k$ of the $k$th sample with the upper and lower bounds of the performance bins. If $T^k$ falls into the $m$th bin by meeting $T_{m,l} < T^k \leq T_{m,u}$, the number of the chips in this bin is increased by one. Instead of comparing $T^k$ with the bounds of the bins directly, we take advantage of the fact that the yield values of the circuit in different bins are a part of the optimization objective defined in (4) and the price of a chip in a high performance bin is higher than that in a low performance bin. We define the 0-1 variables $g_m^k$, $m = 1, \ldots, N_p$ to represent whether the minimum clock period $T^k$ of the $k$th sample is smaller than the upper bound of the $m$th bin. Therefore, $g_m^k$ can be constrained as

$$g_m^k = 1 \iff T^k \leq T_{m,u}, m = 1, 2, \ldots, N_p. \quad (12)$$

We then use $g_m^k$ to define another 0-1 variable $b_m^k$ which indicates whether the $k$th sample falls into the $m$th bin meeting $T_{m,l} < T^k \leq T_{m,u}$, as

$$b_m^k = \begin{cases} g_m^k & m = 1, \\ g_m^k - g_{m-1}^k & m = 2, \ldots, N_p. \end{cases} \quad (13)$$

The constraint (12) can be transformed into

$$T^k - T_{m,u} \leq (1 - g_m^k)\Gamma, m = 1, 2, \ldots, N_p \quad (14)$$

where $\Gamma$ is very large positive constant.

The constraints (13) and (14) can be explained as follows. If $T^k$ is no larger than the upper bound of the $m$th bin $T_{m,u}$, the left side of (14) is negative, so that $g_m^k$ can be either 0 or 1; otherwise, $g_m^k$ must be 0. Since the objective of the optimization problem is to increase the numbers of chips in high-performance bins as much as possible, the solver will assign all $g_m^k, g_{m+1}^k, \ldots, g_{N_p}^k$ to 1 if $T^k \leq T_{m,u}$, because the bins are arranged in the high performance to low performance order so that $T^k$ is also smaller than $T_{m+1,u}, \ldots, T_{N_p,u}$. Therefore, the constraint (13) only keeps the $b_m^k$ for the fastest bin to which the sample can be assigned to be 1, and for the slower bins it is set to 0. Consequently, $b_m^k$ represents whether the chip is assigned to the $m$th bin.

With $b_m^k$ we can calculate the numbers of emulated chips in all bins easily, and the yield or the percentage $y_m$ for the $m$th bin can be expressed as

$$y_m = \sum_{k=1}^{N_s} b_m^k \bigg/ N_s \quad (15)$$

where $N_s$ is the total number of samples.

With the constraints defined above, the problem to optimize the overall profit can be expressed as

$$\text{maximize} \quad \sum_{m=1}^{N_p} p_m y_m \quad (16)$$

$$\text{s.t.} \quad (5)-(7), (9)-(11), (13)-(15),$$
$$\text{w.r.t. all flip-flops pair indexed by } (i, j), \quad (17)$$
$$\text{and } k = 1, \ldots, N_s.$$

The basic idea of this formulation is that we use a given number of samples to emulate chips after manufacturing and model the bin assignment process. We then use an ILP solver to maximize the profit in this simulated scenario to determine which flip-flops should have buffers. Since the relation between the locations of buffers and the yield assignment is established in this formulation, we can determine the locations of buffers directly by solving the optimization problem above. In previous methods [9], [10], the relation between buffer locations and yield is not analyzed directly. Instead, these methods consider this relation as a separate evaluation problem, and the yield values for different combinations of buffer locations are calculated using Monte Carlo simulation, and only used as a metric to determine the next decision points in the path search or cutting plane methods. Consequently, Monte Carlo simulation have to be executed many times, resulting in a large runtime.

If the number of emulated samples $N_s$ in the integer linear optimization problem (16)–(17) is large enough, the profit can be modeled accurately and the values of $c_i$ in the solution indicate the optimal locations to insert tuning buffers for the maximum profit. However, a large $N_s$ may increase the number of constraints in (17) to the degree that the size of the ILP problem exceeds the capacity of all existing ILP solvers. To deal with this scalability problem, we apply two techniques: 1) we reduce the number of emulated samples $N_s$ by using a low-discrepancy sample sequence instead of a purely random sampling sequence; 2) we split the problem (16)–(17) into subsets and use them to learn the locations of buffers iteratively. After each iteration, the candidates of buffer locations can be refined.

*B. Reducing the number of emulation samples using a low-discrepancy sequence*

The sampling-based concept above requires a large number of samples to guarantee the quality of the resulting buffer locations. Consider the extreme case where we use only two samples, which have different probabilities to appear in the manufactured chips. In the formulation (16)–(17), however, we do not differentiate these two samples with respect to their probabilities so that the two samples have the same influence on the selection of buffers. Consequently, the formulation loses accuracy because the calculated optimal profit deviates from the real profit.

In traditional Monte Carlo simulation methods, this discrepancy problem is solved by using a large number of samples. Since the samples are generated according to the joint distribution of the variables, the number of points falling into a part of the sampling space corresponds to the probability of that region. The effect of probability can thus be handled by (16)–(17) implicitly, because samples from regions with large probabilities in the problem space appear more often than samples from other regions. Another way to solve this discrepancy problem is to use the probability of representative samples as further coefficients of the yield values in the objective (16) directly. But it is not clear how many samples should be generated to guarantee the quality of the result.

The third method to solve the problem of a large sampling number is to use a low-discrepancy sequence such as studied in [27]. In such a sequence, the number of samples in a given part of the sampling space is proportional to the probability of that region. The advantage of such a sequence is that this quasi-random sequence ensures the low discrepancy even with a small number of samples, so that it is widely used in quasi-Monte Carlo methods to reduce runtime. In statistical timing analysis, this method also demonstrates a strong advantage, e.g., more than 20 times acceleration has been achieved in [28]. In this paper, we use the Sobol sequence in [29] to reduce the number of samples $N_s$. The effect of this sequence can be demonstrated using the examples in Fig. 5, where Fig. 5a shows a purely random number sequence of 256 samples for two uniform-distributed variables. Fig. 5b demonstrate that the Sobol sequence with the same number of samples spreads more evenly in the space. The original Sobol sequence follows uniform distribution, and it can be transformed to other distributions easily using methods such as the Box-Muller transform [30]. In our method, we use 1000 samples in the Sobol sequence, which are one tenth of the usually used 10,000 samples of random variables in statistical static timing analysis [2]. In practice, test cases can converge even earlier with fewer than 1000 samples.

*C. Buffer allocation with prefiltering and iterative learning*

In the $N_s$ samples, some might be fast enough to be assigned into the fastest bin without tuning; others might be too slow to be tuned into the slowest bin, even with all flop-flops connected with tuning buffers. In both scenarios, tuning buffers play no role in improving the overall profit. Therefore, we exclude these samples from the ILP formulation (16)–(17) to reduce the number of variables and constraints.

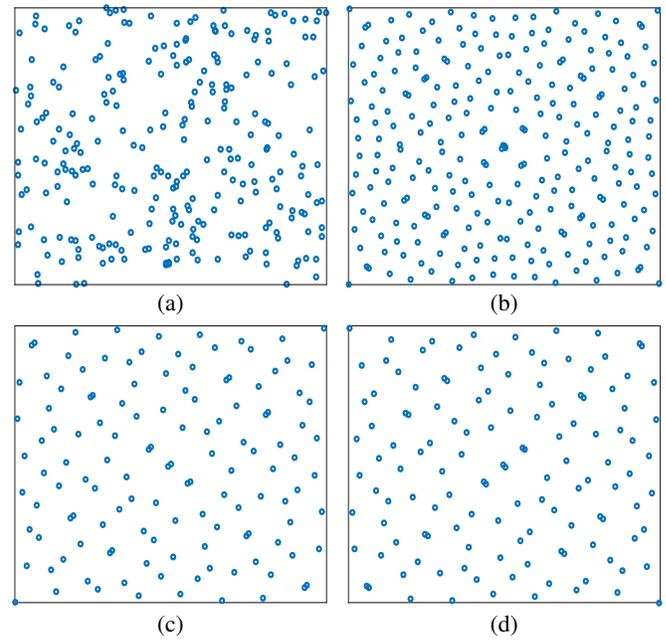

Fig. 5. Purely random sequence and Sobol low-discrepancy sequence. (a) 256 random samples of two uniform variables. (b) 256 samples of a Sobol sequence for two uniform variables. (c) The first 128 samples from the Sobol sequence in (b). (d) The next 128 samples from the Sobol sequence in (b).

To filter out the samples of the first type, we need only to set all values of tuning buffers, $x_i^k$ and $x_j^k$ in (5) and (6) to 0, and calculate the clock period $T_{min}^k$ for this sample as $T_{min}^k = \max_{i,j}\{\overline{d}_{ij}^k + s_j^k\}$. If $T_{min}^k$ is smaller than the upper bound of the fastest bin, this sample is fast enough and no tuning is required. The constraint (6) is checked similarly. If all these constraints can be met without tuning buffers, this sample is excluded.

To filter out the samples that are too slow to be assigned to a bin even with extensive buffer tuning, we evaluate each path delay in a sample by verifying whether it is possible to tune this path to meet the upper bound of the slowest bin without considering the other paths. In the constraint (5), the sum of the path delay $\overline{d}_{ij}^k + s_j^k$ and $x_i^k - x_j^k$ should be no larger than $T^k$. We set buffer values $x_i^k$ and $x_j^k$ to the smallest and the largest values that are possible according to buffer specifications, respectively, and check whether the resulting clock period $T^k$ is smaller than the upper bound of the slowest bin. If this still does not hold, there is no chance that this sample can be assigned to one of the bins and the corresponding sample is not included in the profit optimization problem. We repeat this prefiltering checking using (6) to exclude samples that do not work in any case due to unavoidable hold time violations.

After prefiltering, the remaining samples are used to determine buffer locations by solving the optimization problem (16)–(17). The number of these remaining samples is denoted as $N_f$. For a large circuit, the number of remaining variables and constraints in this ILP problem may still be too large to be dealt with by a modern solver. To reduce the scale of the ILP problem further, we split the ILP problem (16)–(17) into subproblems and determine the buffer locations with an iterative flow based on: 1) a subsequence of a Sobol sequence still exhibits a good low discrepancy as shown in Fig. 5c-d; 2) in a circuit only a small number of buffers can be inserted due to area cost. The iterative flow is illustrated in Fig. 4.

The first fact above shows that we may solve the ILP problem (16)–(17) with only a part of the Sobol sequence, meaning that we can capture the buffer locations only using a subset of samples. Therefore, we partition the whole Sobol sequence into several parts so that each part contains $N_t$ samples which are processed together in one ILP problem (16)–(17). We call the samples processed in one ILP problem a batch. In our implementation, the number of samples $N_t$ in one batch is determined by evaluating the numbers of variables and constraints and the capacity of the ILP solver. Since variables in an ILP problem define the dimension of the problems space, they carry more complexity into the ILP problem than constraints. Therefore, we consider the complexity of a variable to be five times that of a constraint, and the total number of the equivalent constraints should be smaller than a constant, $2 \times 10^6$ for Gurobi [31] used in our experiments.

Though the samples in subsequences generally have lower discrepancy compared with a purely random sequence, there are still some slight patterns in these subsequences because of the small number of samples in one subsequence, as shown in Fig. 5c-d. Consequently, a subsequence with a limited number of samples may not capture all the buffer locations. We alleviate this problem by combining the buffer locations captured by different subsequences into a buffer set $\mathcal{B}'$. Once we finish solving (16)–(17) with all sample batches, the buffer locations in $\mathcal{B}'$ are the possible locations to insert buffers, as shown in the inner loop in Fig. 4. In this loop, we also relax the number of buffers from $N_b$ to $\beta N_b$ in the constraint (11) ($\beta = 1.5$ in our experiments) to increase the coverage of potential buffer locations captured by the subsequences. We will use a group technique to reduce the number of buffers back to $N_b$ after all location candidates are captured. The inner iterative flow stops if no new buffer is added into the buffer set $\mathcal{B}'$ in the past three iterations.

After processing all sample batches in the inner loop, we execute the iterative buffer allocation flow as the outer loop in Fig. 4. In these iterations, only the buffer candidates in $\mathcal{B}$ need to be modeled with variables $c_i$ as in (8) and only the delays of paths connected to these buffer candidates need to be sampled as (5)–(7). Consequently, more samples can be processed in one iteration so that the number of batches $\lceil N_f / N_t \rceil$ can be reduced. With these outer iterations, buffer locations are gradually refined and the outer loop finishes if the number of batches cannot be decreased.

*D. Reducing buffer area by tuning concentration and grouping*

The iterative optimization flow in Fig. 4 only determines the locations to insert buffers for profit improvement after manufacturing. But the sizes of the buffers are not addressed.

In this section, we introduce a method to concentrate tuning values toward each other and to group buffers thereafter.

The concept of area reduction can be explained using Fig. 6. After executing the iterative buffer allocation in Fig. 4, the tuning values of a buffer in all samples may be scattered in a wide range such as in Fig. 6a, because the solver only minimizes the number of buffers, but does not consider the relation between the tuning values of different samples, so that it only returns one of the many feasible tuning combinations. If we can concentrate the tuning values toward each other, the real ranges of the buffers which cover all the tuning values appearing in the samples can be reduced. In addition, the concentrated tuning values may exhibit a high correlation by forming similar trends of tuning values as in Fig. 6c. This resemblance can thus be used to group buffers.

To push the scattered tuning values into a narrower range, we minimize their absolute values in the optimization, as illustrated in Fig. 6a. In this way, the solver tries to return the buffer values around 0 as much as possible using only the buffer candidates in $\mathcal{B}$ and guaranteeing the profit $\mathcal{P}$ calculated by executing the flow in Fig. 4. This process is formulated as follows.

$$\text{minimize} \sum_{i \in I_\mathcal{B}, k} |x_i^k| \qquad (18)$$

$$\text{s.t.} \quad (5)\text{–}(7), (9)\text{–}(11), (13)\text{–}(15),$$
w.r.t. all flip-flops pair indexed by $(i, j)$, (19)
and $k = 1, \ldots N_f$, and

$$\sum_{m=1}^{N_p} p_m y_m \geq \mathcal{P} \qquad (20)$$

where $I_\mathcal{B}$ is the index set of all buffer locations in $\mathcal{B}$. The objective function (18) can be transformed into a linear form easily as explained in [32].

The difference between the optimization problem (18)–(20) and the optimization problem (16)–(17) includes: 1) the objective becomes the sum of the absolute values of all tuning values; 2) the buffer candidates are narrowed as the buffer set $\mathcal{B}$ returned by the flow in Fig. 4; 3) the profit becomes a constraint to guarantee the tuning range concentration does not affect the profit. By solving the problem (18)–(20), all tuning values are pushed toward zero as illustrated in Fig. 6b, so that the buffer ranges become more compact.

Another technique to reduce area cost is to group buffers that have similar tuning patterns into one buffer. For example, if two buffers have very similar tuning values in all samples, only one buffer needs to be built in the circuit and the delayed clock signal is connected to two flip-flops. To make the patterns in buffer tuning more obvious, we first calculate the weighted average of all tuning values of a buffer after solving (18)–(20). Afterwards, the buffer tuning values are pushed further toward this average. This process makes the number of different tuning values smaller, so that it is easier for two buffers to have similar tuning patterns. The result of this step is that buffer tuning values may form a peak at the tuning average as illustrated in Fig. 6c. This step is very similar to the

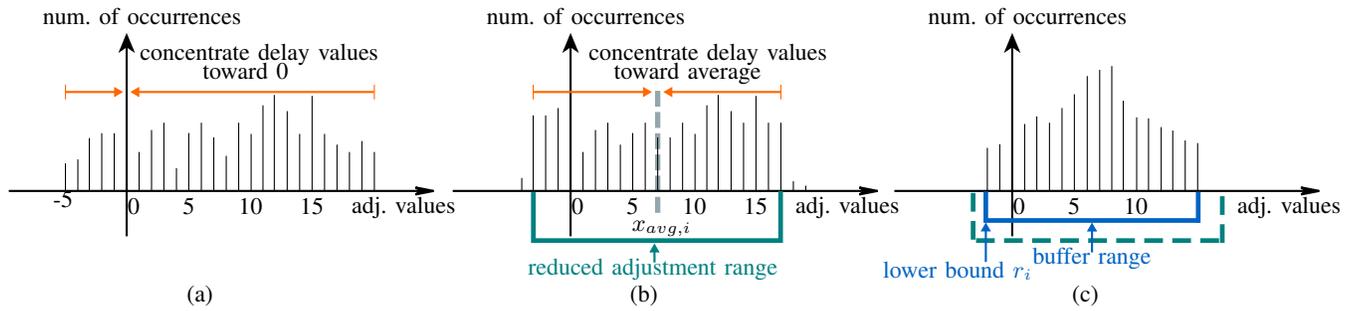

Fig. 6. Concentrating tuning values of a buffer in all samples. The x-axis represents the adjusted delays of the buffer in all samples, and the y-axis the number of occurrences of the discrete delay values. (a) Scattered tuning values. (b) Tuning values concentrated toward zero. (c) Reduced buffer range after concentrating tuning values toward the average.

problem formulation in (18)–(20), except that the optimization objective is replaced by

$$\text{minimize} \sum_{i \in I_{\mathcal{B}}, k} |x_i^k - x_{avg,i}| \quad (21)$$

where $x_{avg,i}$ is the weighted average of all tuning values calculated from the result of solving (18)–(20).

After tuning values are concentrated, we try to cover all the tuning values using the smallest range window. The upper bound of the size of this range window is predefined as $\tau_i$ in (3). As shown in Fig. 6c, the range window slides along the x-axis. Since the y-axis represents the numbers of the corresponding tuning value occurrences in all samples, the total number of buffer tunings covered by the window is the sum of the tuning occurrences in the window. For yield improvement, we select the range window that covers the largest number of tunings, meaning that these tuning values are feasible in post-silicon configuration. The other values that fall out of the window are discarded. With this step, both the buffer size $\tau_i$ and the lower bound $r_i$ in (3) are determined.

In the last step of buffer insertion, we group buffers with similar tuning values to reduce the number of buffers inserted into the circuit. Buffers in the same group are implemented by only one physical buffer and the tuning values are shared by all the flip-flops connected to the buffer. The concept of grouping is illustrated in Fig. 7.

In grouping buffers, we first calculate the correlation coefficients of tuning values of buffer pairs. If the mutual correlation coefficients between several buffers are all above the threshold $r(i, j)$ and their distance is smaller than $d(i, j)$, they are grouped together and implemented with only one physical buffer. In practice, designers can also constrain the total number of buffers in the circuit as $N_b$. If the number of buffers after grouping still exceeds the specified number, the buffers with the fewest tunings are removed until the number of buffers meets the specification.

### E. Acceleration techniques

To improve the efficiency of the proposed method, we sample statistical delays between flip-flops directly instead of sampling delays of combinational gates. For example, the delays in (1) and (2) are calculated using a statistical timing engine only once. We then generate a Sobol sequence from these statistical delays directly, instead of executing a static timing analysis algorithm for each sample.

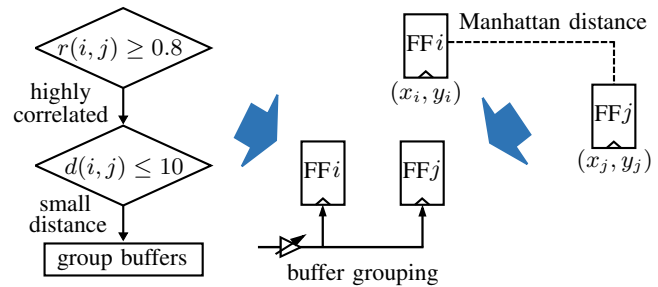

Fig. 7. Buffer grouping according to tuning correlation and distance. Correlation threshold $r(i, j)$ is set to 0.8. Distance threshold $d(i, j)$ between buffers is set to ten times of the minimum distance between flip-flops.

In addition, we filter connections between flip-flops according to their statistical distributions. If the $3\sigma$ delay of a path is still small enough not to affect the circuit performance, this path is not included when creating the constraints (1)–(2). For example, in the constraint (1) we first set $x_i$ to the largest value and $x_j$ to the smallest value in the range windows, respectively, and $\bar{d}_{ij}$ and $s_j$ to their $3\sigma$ values. If this extreme setting still allows this path to work with a clock period in the fastest bin, this path is simply discarded from the problem formulation. Similarly, we also filter hold time constraints (2) according to the -$3\sigma$ values of path delays.

## V. EXPERIMENTAL RESULTS

The proposed method was implemented in C++ and tested using a 3.20 GHz CPU with one thread. We demonstrate the results using circuits from the ISCAS89 benchmark set and from the TAU 2013 variation-aware timing analysis contest. The number of flip-flops and the number of logic gates are shown in the columns $n_s$ and $n_g$ in Table II, respectively.

The benchmark circuits in our experiments were sized using a 45 nm library. We assumed that the maximum allowed buffer ranges were 1/8 of the original clock period and tuning delays of the buffers were discrete with 20 steps, as in [7]. The standard deviations of transistor length, transistor width and oxide thickness were set to 15.7%, 11.1% and 5.3% of the nominal values, respectively. We used Gurobi [31] to solve the optimization problems in the proposed method.

TABLE II
RESULTS OF BUFFER ALLOCATION FOR POST-SILICON BINNING

| Circuit | | | Buffer | | With Buffer Allocation | | | | | | Runtime |
|---|---|---|---|---|---|---|---|---|---|---|---|
| | $n_s$ | $n_g$ | $n_b$ | $s_b$ | $\mathcal{Y}_{b1}$ | $\mathcal{Y}_{b2}$ | $\mathcal{Y}_{b3}$ | $\mathcal{Y}_{\mu_T+\sigma_T}$ | $\mathcal{Y}_{inc}(\%)$ | $\mathcal{P}_{inc}(\%)$ | $t_c(s)$ |
| s9234 | 211 | 5597 | 2 | 18.00 | 52.31% | 18.80% | 13.72% | 84.83% | 0.70% | 3.37% | 20.83 |
| s13207 | 638 | 7951 | 6 | 13.83 | 63.40% | 13.55% | 11.03% | 87.98% | 3.85% | 18.47% | 34.93 |
| s15850 | 534 | 9772 | 5 | 7.20 | 67.93% | 14.01% | 10.16% | 92.10% | 7.97% | 26.18% | 56.81 |
| s38584 | 1426 | 19253 | 14 | 12.52 | 63.79% | 16.33% | 10.71% | 90.83% | 6.70% | 20.62% | 71.03 |
| mem_ctrl | 1065 | 10327 | 10 | 13.06 | 58.41% | 17.49% | 12.93% | 88.83% | 4.70% | 12.76% | 164.62 |
| usb_funct | 1746 | 14381 | 17 | 14.71 | 54.61% | 17.58% | 14.03% | 86.22% | 2.09% | 6.67% | 147.88 |
| ac97_ctrl | 2199 | 9208 | 21 | 13.08 | 57.96% | 16.45% | 12.85% | 87.26% | 3.13% | 11.39% | 115.93 |
| pci_bridge32 | 3321 | 12494 | 33 | 8.08 | 60.02% | 16.84% | 12.00% | 88.86% | 4.73% | 14.87% | 1816.81 |
| | | | Average | 12.56 | 59.80% | 16.38% | 12.18% | 88.36% | 4.23% | 14.29% | |
| | | | Yield without buffers | | 50.00% | 19.15% | 14.98% | 84.13% | | | |

We used three bins in the experiments to improve the overall profit. The boundaries between these bins were set to $\mu_T$, $\mu_T + 0.5\sigma_T$ and $\mu_T + \sigma_T$, where $\mu_T$ and $\sigma_T$ are the mean value and the standard deviation of the clock period of the original circuit without clock buffers. Chips with clock period larger than $\mu_T + \sigma_T$ were considered as yield loss. With this setting, the original yield values of these three bins without tuning buffers are 50%, 19.15%, and 14.98%, respectively. In all these test cases, the numbers of allocated buffers $N_b$ were constrained as lower than 1% of the numbers of flip-flops in the circuits, as shown in the $n_b$ column. After allocating post-silicon tuning buffers using the proposed method, we ran Monte Carlo simulation with these circuits to verify the yield improvement. In the simulation, we generated 10 000 samples. For each sample we calculated its minimum clock period using an ILP solver due to the appearance of tuning buffers, and assigned the sample to one of the performance bins. The yield value of a circuit in a bin is the number of samples in that bin divided by 10 000. The samples in our experiments are conceptually different from the samples discussed in Section IV, because they were only used to emulate post-silicon measurements. For each sample, we verify whether a chip can be assigned into a bin by solving the classical skew scheduling problem in [19]. In reality, the delays and timing properties cannot be measured exactly from the manufactured chips, so that the actual yield is slightly smaller than the reported yield, as discussed in [21]. This yield, however, still serves as a good indicator to determine buffer locations.

The yield values of the three bins are shown in the columns $\mathcal{Y}_{b1}$, $\mathcal{Y}_{b2}$ and $\mathcal{Y}_{b3}$ in Table II, respectively. Compared with the yield values without clock buffers, we can see that the yield in the first bin is increased significantly but the yield values of the other two bins are smaller, because with tuning buffers chips have a better chance to be tuned to a higher performance. Adding the yield values of the three bins together, we can calculate the yield of a circuit with respect to $\mu_T + \sigma_T$, shown in the $\mathcal{Y}_{\mu_T+\sigma_T}$ column. Compared with the original yield 84.13%, the yield increase is shown in the column $\mathcal{Y}_{inc}$, with an average 4.23%.

With these yield values in the three bins, we can calculate the profit using (4). In the experiments, we set the profit per chip of the three bins to 6, 2, and 1, respectively. The overall profit increase is shown in the column $\mathcal{P}_{inc}$, with an average 14.29%. If we compare the column $\mathcal{P}_{inc}$ and the column $\mathcal{Y}_{inc}$, we can see that the improvement of profit is much more significant than the overall yield improvement due to the introduced tuning buffers and clock binning. To achieve this profit improvement, the number of buffers in the circuit is still less than 1% of the number of flip-flops. If we assume that a buffer takes 10 times area of a flip-flop and flip-flops take 5% of the die area, the area cost of these buffers is about 0.5% of the die area. Therefore, we can expect a good overall revenue improvement, even when we consider the potential cost of post-silicon configuration. A concrete evaluation of this cost will be our future work.

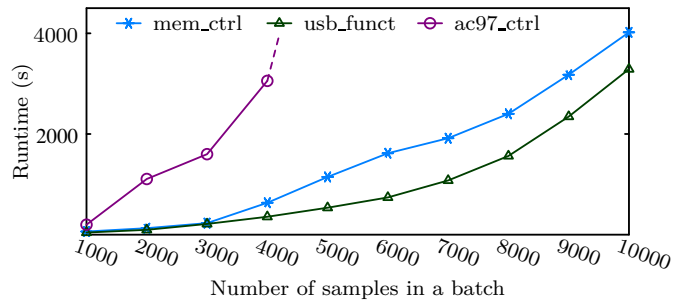

Fig. 8. Scalability trend of the proposed method with a fixed number of samples in each batch.

In the proposed method, we also reduced the buffer sizes by concentrating tuning values. The average buffer sizes in the benchmark circuits are shown in the column $s_b$. Compared with the maximum allowed size 20, the buffer sizes have been reduced effectively by the proposed method while maintaining a good profit improvement. The execution time of the proposed method is shown in the last column of Table II. The largest execution time of the proposed method is 1816.81 seconds, which is already acceptable because the proposed method is executed offline only for a few times.

Since the runtime of solving an ILP problem depends on the structure of constraints as well as their relations, it is difficult to analyze the scalability of the proposed method theoretically. Instead, we tested this method by fixing the number of samples in each batch to solve the buffer insertion problem with respect to a given clock period $\mu_T + \sigma_T$ as used in Table II. The relation between the number of samples in a batch and the runtime is illustrated in Fig. 8. *pci_bridge32* did not finish due to memory limitation, so that it was not included in this evaluation. According to these results, the runtime increases exponentially with respect to the number of samples, especially with large circuits. In the proposed method, the number of samples in each batch is limited to $N_t$ as discussed in Section IV-C. This limitation might lead to a yield degradation because the optimization problem is split into several small problems. To verify the quality of the results produced by the proposed method, we compared them with the yield results of a brute-force method processing 10 000 samples as a whole, as shown in Fig. 9a. With this comparison, it can be observed that the yield degradation of the proposed method is negligible, because the proposed work flow in Fig. 4 first tries to capture all the buffer locations that have a potential to affect the yield. Afterwards, only these locations are considered in further iterations so that a batch can contain more samples, still leading to a good yield result. The runtime of the brute-force method, however, is much larger than the proposed method, as shown in Fig. 9b.

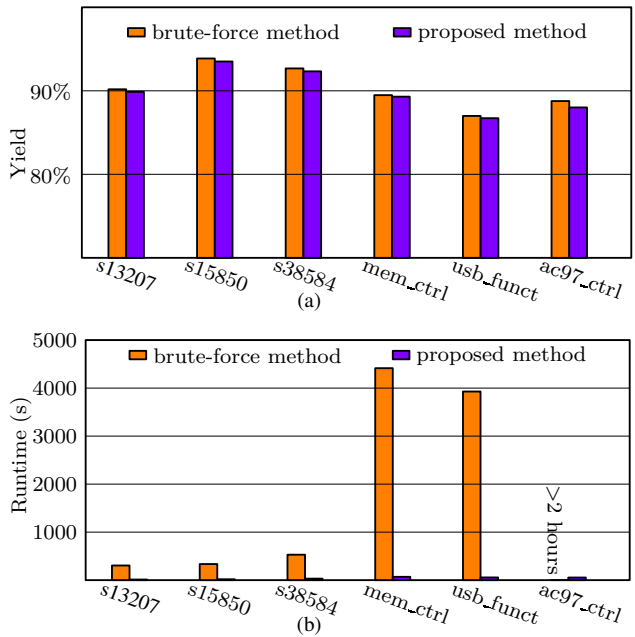

Fig. 9. Yield and runtime comparison between the proposed method and the brute-force method with 10 000 samples. (a) Yield comparison. (b) Runtime comparison.

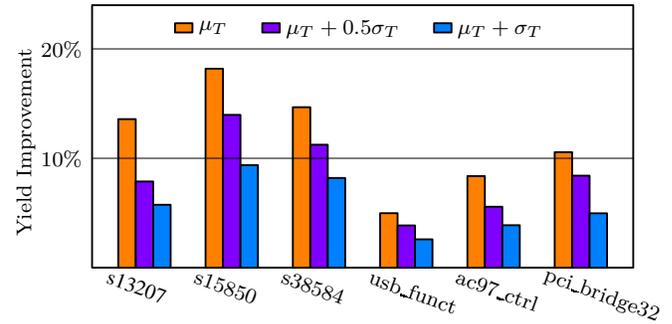

Fig. 10. Yield improvement with clock tuning buffers with respect to $\mu_T$, $\mu_T + 0.5\sigma_T$ and $\mu_T + \sigma_T$, compared with the yield values without tuning buffers.

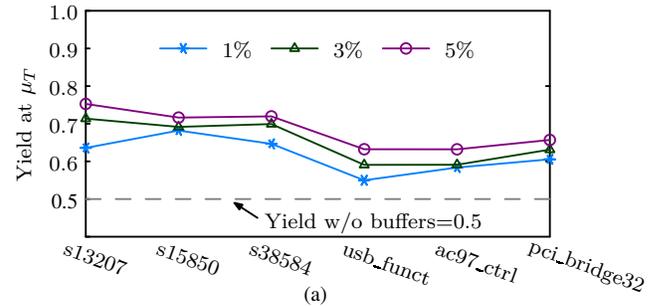

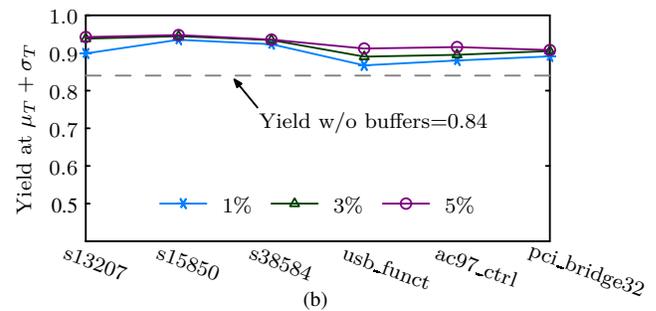

Fig. 11. Yield increase with respect to different numbers of tuning buffers. (a) The target clock period is set to $\mu_T$. (b) The target clock period is set to $\mu_T + \sigma_T$.

In the profit definition (4), if we use only one bin, the problem formulation becomes the problem to improve the yield with respect to a single clock period. In our experiments, we tested this single-bin setting using $\mu_T$, $\mu_T + 0.5\sigma_T$, and $\mu_T + \sigma_T$ as the upper bounds of the single bins, respectively. The results of yield improvement are shown in Fig. 10. In all these test cases, the yield values have been improved effectively, up to 18.19% for the circuit s15850 in the $\mu_T$ bin. In these test cases, the yield improvement is consistently better for bins with higher performance, because in these bins the original yield values without tuning buffers are lower so that there is a large potential for the tuning buffers to take effect.

In our experiments, we constrained the number of buffers to be smaller than 1% of the number of the flip-flops. If this number can be increased, we can expect an increase of yield because there are more chances to tune the chips after manufacturing. To show the effect of more tuning buffers, we tested the numbers of buffers equal to 1%, 3%, and 5% of the number of flip-flops. For each of these buffer numbers, we calculated the yield values with respect to the single clock periods $\mu_T$ and $\mu_T + \sigma_T$, respectively. The results are shown in Fig. 11. According to these experiments, we can see that the yield generally increases when the number of buffers inserted into the circuit increases. Similar to the trend of the yield improvement with respect to different clock periods in Fig. 10, the yield improvement with respect to $\mu_T$ in Fig. 11a is more obvious compared with the yield improvement with respect to $\mu_T + \sigma_T$ in Fig. 11b. For the former, the average improvement of the 5% setting to the 1% setting is 6.78%, but for the latter this improvement is only 2.75%. Consequently, we can conclude that post-silicon buffers are more useful in high-performance designs, specially with clock binning, where the potential for profit/yield improvement is large.

TABLE III
RUNTIME COMPARISON W/O AND W/ ACCELERATION TECHNIQUES

| Circuit | s15850 | s38584 | ac97_ctrl | pci_bridge32 |
|---|---|---|---|---|
| Without acceleration (s) | 3411.29 | 8435.14 | 15967.7 | > 8h |
| With acceleration (s) | 56.81 | 71.03 | 115.9 | 1816.811 |

TABLE IV
YIELD COMPARISON WITH [9]

| Circuit | $N_1$ | $Y_1$ | $N_2$ | $Y_2$ |
|---|---|---|---|---|
| s9234 | 8 | 96.94% | 2 | 98.57% |
| s13207 | 18 | 98.95% | 6 | 99.40% |
| s15850 | 21 | 99.24% | 5 | 99.96% |
| s38584 | 162 | 98.17% | 14 | 99.70% |

TABLE V
YIELD COMPARISON OF DIFFERENT GROUPING ALGORITHMS

| Circuit | s15850 | s38584 | ac97_ctrl | pci_bridge32 |
|---|---|---|---|---|
| $Y_1$ | 84.57% | 85.43% | 84.67% | 84.16% |
| $Y_2$ | 93.51% | 92.33% | 88.01% | 89.10% |

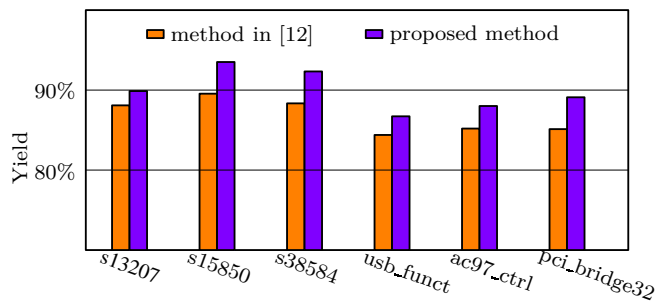

Fig. 12. Yield comparison with the buffer insertion method in [12].

To reduce the execution time of the proposed method, we introduced several acceleration techniques. With the Sobol sequence, the inner loop of the iterative flow in Fig. 4 converged with the test cases *usb_funct* and *pci_bridge32*, while the other cases used up all the samples. To demonstrate the efficiency of the acceleration techniques, we disable all of them and show the execution time in Table III. According to this comparison, it is obvious that the proposed acceleration techniques can shorten the execution time effectively.

The buffer insertion problem is also addressed in [9] with a direct statistical model. For comparison, we show the results from their paper and the results of our method applied to the same set of circuits in Table IV. The $N_1$ column shows the number of buffers in [9], and the $N_2$ column that of our method. Note their method is designed for a clock network with a tree structure and they do not group buffers as we do. Consequently, there is a large difference between the numbers of buffers. The columns $Y_1$ and $Y_2$ show the yield values from their method and our method with the same clock period setting. In this comparison, the proposed method outperforms the method in [9] with a higher yield, while the number of clock tuning buffers is much smaller. Furthermore, we have implemented the method in [12] and the yield comparison is shown in Fig. 12. In this comparison, the numbers of inserted buffers are equal, so that we can conclude that the proposed method outperforms the method in [12] consistently.

In the last step of the proposed method, we group buffers according to the correlation between tuning values. This correlation information is a natural result of the sampling-based method. In [12], a grouping algorithm is also proposed according to circuit structure and distances between flip-flops. We compare the results of our correlation-based grouping method with theirs and the results are shown in Table V, where $Y_1$ is the yield with the grouping algorithm in [12] and $Y_2$ is the yield with the proposed correlation-based grouping. For comparison, we have changed the numbers of buffers in the proposed method so that they are equal to the ones in [12]. From this comparison, we can see that our method produces a better yield, because we have the correlation information from emulated samples.

The method proposed in [1] uses the same concept in this paper, but it captures the locations of buffers by processing emulated samples once at a time. Therefore, the relation between tuning values in different samples is not incorporated. In addition, the method in [1] uses a purely random sequence so that the number of samples is still large. To verify the improvement of the proposed method, we mapped the circuits used in [1] to the same library and tested the yield improvement with respect to $\mu_T$. The results are shown in Fig. 13a, where we can see that the proposed method produces a better yield improvement than [1] with the same number of buffers. Furthermore, we show the execution time of these methods in Fig. 13b. It is clear that the extended method in this paper is more efficient than [1].

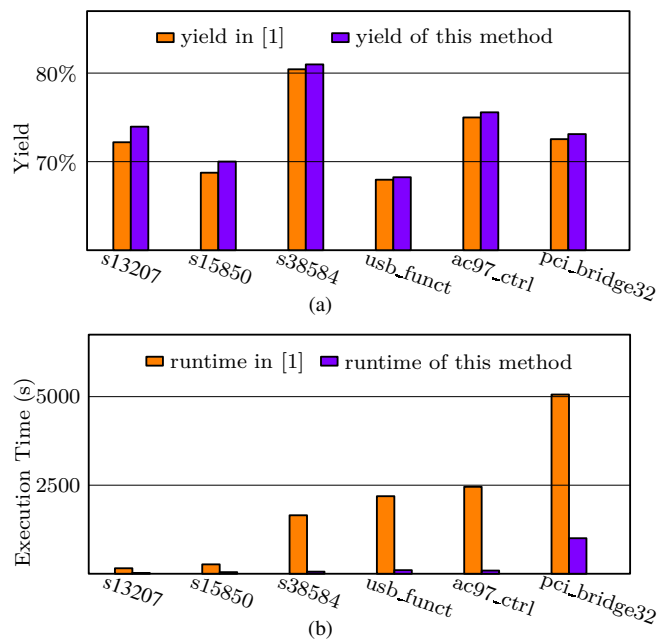

Fig. 13. Comparison with [1]. (a) Yield improvement with the same setting. (b) Comparison of execution time of both methods.

## VI. CONCLUSION

In this paper, we propose a sampling-based method to determine locations and ranges of post-silicon tuning buffers in a circuit to improve the overall profit with clock binning. By establishing the relation between buffer locations and the yield with an ILP model directly, the proposed method can learn the buffer locations for yield improvement effectively. With acceleration techniques such as a low discrepancy sequence, the proposed method takes much less time than previous methods. Experimental results confirm that the profit of the circuit after manufacturing can be improved significantly with a small number of buffers. Future tasks of this work include post-silicon testing and configuration of delays buffers to achieve the given clock period or profit. The major challenge is to make a good tradeoff between test cost and profit improvement.

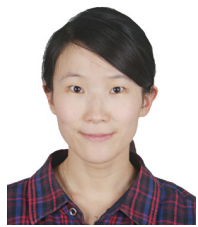

**Grace Li Zhang** received the master's degree from the school of microelectronics, Xidian University, Xi'an, China, in 2014. She is currently pursuing the Ph.D. degree with the Institute for Electronic Design Automation, Technical University of Munich (TUM). Her research interests include high-performance and lower-power design, as well as emerging systems.

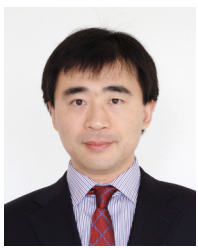

**Bing Li** received the bachelor's and master's degrees in communication and information engineering from Beijing University of Posts and Telecommunications, Beijing, China, in 2000 and 2003, respectively, and the Dr.-Ing. degree in electrical engineering from Technical University of Munich (TUM), Munich, Germany, in 2010. He is currently a researcher with the Institute for Electronic Design Automation, TUM. His research interests include high-performance and lower-power design, as well as emerging systems.

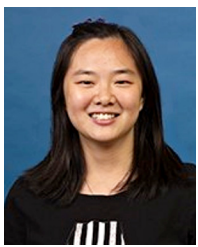

**Jinglan Liu** received her B.E. degree in Communication Engineering from Beijing University of Posts and Telecommunications, Beijing, in 2014. She is currently pursuing the Ph.D. degree in Department of Computer Science and Engineering at University of Notre Dame, Notre Dame, Indiana. Her research interests focus on low-power system design, machine learning applications on interdisciplinary fields.

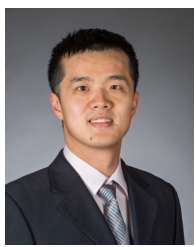

**Yiyu Shi** (SM'06) is currently an associate professor in the Departments of Computer Science and Engineering and Electrical Engineering at the University of Notre Dame. He received his B.S. degree (with honors) in Electronic Engineering from Tsinghua University, Beijing, China in 2005, the M.S and Ph.D. degree in Electrical Engineering from the University of California, Los Angeles in 2007 and 2009 respectively. His current research interests include three-dimensional integrated circuits, and machine learning on chips. In recognition of his research, he has received many best paper nominations in top conferences. He was also the recipient of IBM Invention Achievement Award in 2009, Japan Society for the Promotion of Science (JSPS) Faculty Invitation Fellowship, Humboldt Research Fellowship for Experienced Researchers, IEEE St. Louis Section Outstanding Educator Award, Academy of Science (St. Louis) Innovation Award, Missouri S&T Faculty Excellence Award, National Science Foundation CAREER Award, IEEE Region 5 Outstanding Individual Achievement Award, and the Air Force Summer Faculty Fellowship.

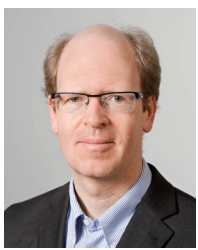

**Ulf Schlichtmann** (S'88–M'90) received the Dipl.-Ing. and Dr.-Ing. degrees in electrical engineering and information technology from Technical University of Munich (TUM), Munich, Germany, in 1990 and 1995, respectively. He was with Siemens AG, Munich, and Infineon Technologies AG, Munich, from 1994 to 2003, where he held various technical and management positions in design automation, design libraries, IP reuse, and product development. He has been a Professor and the Head of the Institute for Electronic Design Automation with TUM, since 2003. He served as the Dean of the Department of Electrical and Computer Engineering, TUM, from 2008 to 2011. His current research interests include computer-aided design of electronic circuits and systems, with an emphasis on designing reliable and robust systems.